# Five Disruptive Technology Directions for 5G

Federico Boccardi, Bell Labs, Alcatel-Lucent
Robert W. Heath Jr., The University of Texas at Austin
Angel Lozano, Universitat Pompeu Fabra
Thomas L. Marzetta, Bell Labs, Alcatel-Lucent
Petar Popovski, Aalborg University

**ABSTRACT**

New research directions will lead to fundamental changes in the design of future 5th generation (5G) cellular networks. This paper describes five technologies that could lead to both architectural and component disruptive design changes: device-centric architectures, millimeter Wave, Massive-MIMO, smarter devices, and native support to machine-2-machine. The key ideas for each technology are described, along with their potential impact on 5G and the research challenges that remain.

## I. INTRODUCTION:

5G is coming. What technologies will define it? Will 5G be just an evolution of 4G, or will emerging technologies cause a disruption requiring a wholesale rethinking of entrenched cellular principles? This paper focuses on potential disruptive technologies and their implications for 5G.

We classify the impact of new technologies, leveraging the Henderson-Clark model [1], as follows:

- Minor changes at both the node and the architectural level, e.g., the introduction of codebooks and signaling support for a higher number of antennas. We refer to these as *evolutions* in the design.
- Disruptive changes in the design of a class of network nodes, e.g., the introduction of a new waveform. We refer to these as *component changes*.
- Disruptive changes in the system architecture, e.g., the introduction of new types of nodes or new functions in existing ones. We refer to these as *architectural changes*.
- Disruptive changes that have an impact at both the node and the architecture levels. We refer to these as *radical changes*.

We focus on disruptive (component, architectural or radical) technologies, driven by our belief that the extremely higher aggregate data rates and the much lower latencies required by 5G cannot be achieved with a mere evolution of the status quo. We believe that the following five potentially disruptive technologies could lead to both architectural and component design changes, as classified in Figure 1.

1. **Device-centric architectures**.
   The base-station-centric architecture of cellular systems may change in 5G. It may be time to reconsider the concepts of uplink and downlink, as well as control and data channels, to better route information flows with different priorities and purposes towards different sets of nodes within the network. We present device-centric architectures in Section II.

2. **Millimeter Wave (mmWave)**.
   While spectrum has become scarce at microwave frequencies, it is plentiful in the mmWave realm. Such a spectrum 'el Dorado' has led to a mmWave 'gold rush' in which researchers with diverse backgrounds are studying different aspects of mmWave transmission. Although far from fully understood, mmWave technologies have already been standardized for short-range services (IEEE 802.11ad) and deployed for niche applications such as small-cell backhaul. In Section III, we discuss the potential of mmWave for a broader application in 5G.

3. **Massive-MIMO**.

Massive-MIMO[1] proposes utilizing a very high number of antennas to multiplex messages for several devices on each time-frequency resource, focusing the radiated energy towards the intended directions while minimizing intra- and inter-cell interference. Massive-MIMO may require major architectural changes, in particular in the design of macro base stations, and it may also lead to new types of deployments. We discuss massive-MIMO in Section IV.

4. **Smarter devices**.
   2G-3G-4G cellular networks were built under the design premise of having complete control at the infrastructure side. We argue that 5G systems should drop this design assumption and exploit intelligence at the device side within different layers of the protocol stack, e.g., by allowing Device-to-Device (D2D) connectivity or by exploiting smart caching at the mobile side. While this design philosophy mainly requires a change at the node level (component change), it has also implications at the architectural level. We argue for smarter devices in Section V.

5. **Native support for Machine-to-Machine (M2M) communication**
   A native[2] inclusion of M2M communication in 5G involves satisfying three fundamentally different requirements associated to different classes of low-data-rate services: support of a massive number of low-rate devices, sustainment of a minimal data rate in virtually all circumstances, and very-low-latency data transfer. Addressing these requirements in 5G requires new methods and ideas at both the component and architectural level, and such is the focus of Section VI.

## II. DEVICE-CENTRIC ARCHITECTURES

Cellular designs have historically relied on the axiomatic role of 'cells' as fundamental units within the radio access network. Under such a design postulate, a device obtains service by establishing a downlink and an uplink connection, carrying both control and data traffic, with the base station commanding the cell where the device is located. Over the last few years, different trends have been pointing to a disruption of this cell-centric structure:

- The base-station density is increasing rapidly, driven by the rise of heterogeneous networks. While heterogeneous networks were already standardized in 4G, the architecture was not *natively* designed to support them. Network densification could require some major changes in 5G. The deployment of base stations with vastly different transmit powers and coverage areas, for instance, calls for a decoupling of downlink and uplink in a way that allows for the corresponding information to flow through different sets of nodes [5].
- The need for additional spectrum will inevitably lead to the coexistence of frequency bands with radically different propagation characteristics within the same system. In this context, [6] proposes the concept of a 'phantom cell' where the data and control planes are separated: the control information is sent by high-power nodes at microwave frequencies whereas the payload data is conveyed by low-power nodes at mm-Wave frequencies. (cf. Section III.)
- A new concept termed *centralized baseband* related to the concept of cloud radio access networks is emerging (cf. [7]), where virtualization leads to a decoupling between a node and the hardware allocated to handle the processing associated with this node. Hardware resources in a pool, for instance, could be dynamically allocated to different nodes depending on metrics defined by the network operator.
- Emerging service classes, described in Section VI, could require a complete redefinition of the architecture. Current works are looking at architectural designs ranging from centralization or

---

[1] Most work in Massive MIMO has assumed operation at frequencies of 5 GHz or less. While the same principles may prove useful at millimeter wavelengths, a successful marriage of Massive MIMO and millimeter waves may take on a considerably different form.

[2] As was learned with MIMO, firstly introduced in 3G as an add-on and then natively included in 4G, major improvements come from a native support, i.e., from a design that is optimized from its inception rather than amended a-posteriori.

- partial centralization (e.g., via aggregators) to full distribution (e.g., via compressed sensing and/or multihop).
  - Cooperative communications paradigms such as CoMP or relaying, which despite falling short of their initial hype are nonetheless beneficial [8], could require a redefinition of the functions of the different nodes. In the context of relaying, for instance, recent developments in wireless network coding [9] suggest transmission principles that would allow recovering some of the losses associated with half-duplex relays. Moreover, recent research points to the plausibility of full-duplex nodes for short-range communication in a not-so-distant future.
  - The use of smarter devices (cf. Section V) could impact the radio access network. In particular, both D2D and smart caching call for an architectural redefinition where the center of gravity moves from the network core to the periphery (devices, local wireless proxies, relays).

Based on these trends, our vision is that the cell-centric architecture should evolve into a device-centric one: a given device (human or machine) should be able to communicate by exchanging multiple information flows through several possible sets of heterogeneous nodes. In other words, the set of network nodes providing connectivity to a given device and the functions of these nodes in a particular communication session should be tailored to that specific device and session. Under this vision, the concepts of uplink/downlink and control/data channel should be rethought (cf. Figure 2).

While the need for a disruptive change in architectural design appears clear, major research efforts are still needed to transform the resulting vision into a coherent and realistic proposition. Since the history of innovations (cf. [1]) indicates that architectural changes are often the drivers of major technological discontinuities, we believe that the trends above might have a major influence on the development of 5G.

### III. MILLIMETER WAVE COMMUNICATION

Microwave cellular systems have precious little spectrum: around 600 MHz are currently in use, divided among operators [10]. There are two ways to gain access to more microwave spectrum:

- To repurpose or *refarm* spectrum. This has occurred worldwide with the repurposing of terrestrial TV spectrum for applications such as rural broadband access. Unfortunately, repurposing has not freed up that much spectrum, only about 80 MHz, and at a high cost associated with moving the incumbents.
- To share spectrum utilizing, for instance, cognitive radio techniques. The high hopes initially placed on cognitive radio have been dampened by the fact that an incumbent not fully willing to cooperate is a major obstacle to spectrum efficiency for secondary users.

Altogether, it appears that a doubling of the current cellular bandwidth is the best-case scenario at microwave frequencies. Alternatively, there is an enormous amount of spectrum at mmWave frequencies ranging from 3 to 300 GHz. Many bands therein seem promising, including most immediately the local multipoint distribution service at 28-30 GHz, the license-free band at 60 GHz, and the E-band at 71-76 GHz, 81-86 GHz and 92-95 GHz. Foreseeably, several tens of GHz could become available for 5G, offering well over an order-of-magnitude increase over what is available at present. Needless to say, work needs to be done on spectrum policy to render these bands available for mobile cellular.

Propagation is not an insurmountable challenge. Recent measurements indicate similar general characteristics as at microwave frequencies, including distance-dependent pathloss and the possibility of non-line-of-sight communication. A main difference between microwave and mmWave frequencies is the sensitivity to blockages: the results in [11], for instance, indicate a pathloss exponent of 2 for line-of-sight propagation but 4 (plus an additional power loss) for non-line-of-sight. MmWave cellular research will need to incorporate sensitivity to blockages and more complex channel models into the analysis, and also study the effects of enablers such as higher density infrastructure and relays. Another enabler is the separation between control and data planes, already mentioned in Section II.

Antenna arrays are a key feature in mmWave systems. Large arrays can be used to keep the antenna aperture constant, eliminating the frequency dependence of pathloss relative to omnidirectional antennas (when utilized at one side of the link) and providing a net array gain to counter the larger thermal noise bandwidth (when utilized at both sides of the link). Adaptive arrays with narrow beams also reduce the impact of interference, meaning that mmWave systems could more often operate in noise-limited rather than interference-limited conditions. Since meaningful communication might only happen under sufficient array gain, new random access protocols are needed that work when transmitters can only emit in certain directions and receivers can only receive from certain directions. Adaptive array processing algorithms are required that can adapt quickly when beams are blocked by people or when some device antennas become obscured by the user's own body.

MmWave systems also have distinct hardware constraints. A major one comes from the high power consumption of mixed signal components, chiefly the analog-to-digital (ADC) and digital-to-analog converters (DAC). Thus, the conventional microwave architecture where every antenna is connected to a high-rate ADC/DAC is unlikely to be applicable to mmWave without a huge leap forward in semiconductor technology. One alternative is a hybrid architecture where beamforming is performed in analog at RF and multiple sets of beamformers are connected to a small number of ADCs or DACS; in this alternative, signal processing algorithms are needed to steer the analog beamforming weights. Another alternative is to connect each RF chain to a 1-bit ADC/DAC, with very low power requirements; in this case, the beamforming would be performed digitally but on very noisy data. There are abundant research challenges in optimizing different transceiver strategies, analyzing their capacity, incorporating multiuser capabilities, and leveraging channel features such as sparsity.

A data rate comparison between technologies is provided in Fig. 3, for certain simulation settings, in terms of mean and 5% outage rates. MmWave operation is seen to provide very high rates compared to two different microwave systems. The gains exceed the 10x spectrum increase because of the enhanced signal power and reduced interference thanks to directional beamforming at both transmitter and receiver.

From the discussion above, and referring again to the Henderson-Clark model, we conclude that mmWave requires radical changes in the system, as it has a strong impact in both the component and the architecture designs. Consequently, we view mmWave as a potentially disruptive technology for 5G, which, provided the afore-discussed challenges can be tackled, could lead to unrivaled data rates and a completely different user experience.

## IV.    MASSIVE MIMO

Massive MIMO (also referred to as 'Large-Scale MIMO' or 'Large-Scale Antenna Systems') is a form of multiuser MIMO in which the number of antennas at the base station is much larger than the number of devices per signaling resource [14]. Having many more base station antennas than devices renders the channels to the different devices quasi-orthogonal and very simple spatial multiplexing/de-multiplexing procedures quasi-optimal. The favorable action of the law of large numbers smoothens out frequency dependencies in the channel and, altogether, huge gains in spectral efficiency can be attained (cf. Fig. 4).

In the context of the Henderson-Clark framework, we argue that massive-MIMO has a disruptive potential for 5G:

- At a node level, it is a scalable technology. This is in contrast with 4G, which, in many respects, is not scalable: further sectorization therein is not feasible because of (*i*) the limited space for bulky azimuthally-directive antennas, and (*ii*) the inevitable angle spread of the propagation; in turn, single-user MIMO is constrained by the limited number of antennas that can fit in certain mobile devices. In contrast, there is almost no limit on the number of base station antennas in massive-MIMO provided that time-division duplexing is employed to enable channel estimation through uplink pilots.
- It enables new deployments and architectures. While one can envision direct replacement of macro base stations with arrays of low-gain resonant antennas, other deployments are possible, e.g., conformal arrays on the facades of skyscrapers or arrays on the faces of water tanks in rural locations. Moreover, the same massive-MIMO principles that govern the use of collocated arrays of antennas apply also to distributed deployments in which a college campus or an entire city could be covered with a multitude of distributed antennas that collectively serve many users (in this framework, the centralized baseband concept presented in Section II is an important architectural enabler).

While very promising, massive-MIMO still presents a number of research challenges. Channel estimation is critical and currently it represents the main source of limitations. User motion imposes a finite coherence interval during which channel knowledge must be acquired and utilized, and consequently there is a finite number of orthogonal pilot sequences that can be assigned to the devices. Reuse of pilot sequences causes pilot contamination and coherent interference, which grows with the number of antennas as fast as the desired signals. The mitigation of pilot contamination is an active research topic. Also, there is still much to be learned about massive-MIMO propagation, although experiments thus far support the hypothesis of channel quasi-orthogonality. From an implementation perspective, massive-MIMO can potentially be realized with modular low-cost low-power hardware with each antenna functioning semi-autonomously, but a considerable development effort is still required to demonstrate the cost-effectiveness of this solution. Note that, at the microwave frequencies considered in this section, the cost and the energy consumption of ADCs/DACs are sensibly lower than at mmWave frequencies (cf. Section III).

From the discussion above, we conclude that the adoption of massive-MIMO for 5G could represent a major leap with respect to today's state-of-the-art in system and component design. To justify these major changes, massive-MIMO proponents should further work on solving the challenges emphasized above and on showing realistic performance improvements by means of theoretical studies, simulation campaigns, and testbed experiments.

## V. SMARTER DEVICES

Earlier generations of cellular systems were built on the design premise of having complete control at the infrastructure side. In this section, we discuss some of the possibilities that can be unleashed by allowing the devices to play a more active role and, thereafter, how 5G's design should account for an increase in device smartness. We focus on three different examples of technologies that could be incorporated into smarter devices, namely D2D, local caching, and advanced interference rejection.

### V.1 D2D

In voice-centric systems it was implicitly accepted that two parties willing to establish a call would *not* be in close proximity. In the age of data, this premise might no longer hold, and it could be common to have situations where several co-located devices would like to wirelessly share content (e.g., digital pictures) or interact (e.g., video gaming or social networking). Handling these communication scenarios via simply connecting through the network involves gross inefficiencies at various levels:

- Multiple wireless hops are utilized to achieve what requires, fundamentally, a single hop. This entails a multifold waste of signaling resources, and also a higher latency.
- Transmit powers of a fraction of a Watt (in the uplink) and several Watts (in the downlink) are consumed to achieve what requires, fundamentally, a few milliWatts. This, in turn, entails unnecessary levels of battery drain and of interference to all other devices occupying the same signaling resources elsewhere.
- Given that the pathlosses to possibly distant base stations are much stronger than the direct-link ones, the corresponding spectral efficiencies are also lower.

While it is clear that D2D has the potential of handling local communication more efficiently, local high-data-rate exchanges could also be handled by other radio access technologies such as Bluetooth or Wi-Fi direct. Use cases requiring a mixture of local and nonlocal content or a mixture of low-latency and high-data-rate constraints (e.g., interaction between users via augmented reality), could represent more compelling reasons for the use of D2D. In particular, we envision D2D as an important enabler for applications requiring low-latency[3], especially in future network deployments utilizing baseband centralization and radio virtualization (cf. Section I).

From a research perspective, D2D communication presents relevant challenges:

- Quantification of the real opportunities for D2D. How often does local communication occur? What is the main use case for D2D: fast local exchanges, low-latency applications or energy saving?
- Integration of a D2D mode with the uplink/downlink duplexing structure.
- Design of D2D-enabled devices, from both a hardware and a protocol perspective, by providing the needed flexibility at both the PHY and MAC layers.
- Assessing the true net gains associated with having a D2D mode, accounting for possible extra overheads for control and channel estimation.

Finally, note that, while D2D is already being studied in 3GPP as a 4G add-on[2], the main focus of current studies is proximity detection for public safety [15]. What we discussed here is having a D2D dimension *natively* supported in 5G.

**V.2 Local Caching**

The current paradigm of cloud computing is the result of a progressive shift in the balance between data storage and data transfer: information is stored and processed wherever it is most convenient and inexpensive because the marginal cost of transferring it has become negligible, at least on wireline networks [2]. For wireless devices though, this cost is not always negligible. The understanding that mobile users are subject to sporadic 'abundance' of connectivity amidst stretches of 'deprivation' is hardly new, and the natural idea of opportunistically leveraging the former to alleviate the latter has been entertained since the 1990s [3]. However, this idea of caching massive amounts of data at the edge of the wireline network, right before the wireless hop, only applies to delay-tolerant traffic and thus it made little sense in voice-centric systems. Caching might finally make sense now, in data-centric systems [4].

Thinking ahead, it is easy to envision mobile devices with truly vast amounts of memory. Under this assumption, and given that a substantial share of the data that circulates wirelessly corresponds to the most popular audio/video/social content that is in vogue at a given time, it is clearly inefficient to transmit such content via unicast and yet it is frustratingly impossible to resort to multicast because the demand is asynchronous. We hence see local caching as an important alternative, both at the radio access network edge (e.g., at small cells) and at the mobile devices, also thanks to enablers such as mmWave and D2D.

**V.3 Advanced Interference Rejection**

In addition to D2D capabilities and massive volumes of memory, future mobile devices may also have varying form factors. In some instances, the devices might accommodate several antennas with the consequent opportunity for active interference rejection therein, along with beamforming and spatial multiplexing. A joint design of transmitter and receiver processing, and proper control and pilot signals, are critical to allow advanced interference rejection. As an example, in Fig. 5 we show the gains obtained by incorporating the effects of nonlinear, intra and inter-cluster interference awareness into devices with 1, 2 and 4 antennas.

While this section has been mainly focused on analyzing the implications of smarter devices at a component level, in Section II we discussed the impact at the radio access network architecture level. We regard smarter devices as having all the characteristic of a disruptive technology (cf. Section I) for 5G, and therefore we encourage researchers to further explore this direction.

---

[3] Low-latency local communications are also discussed in Section VI. However, while the focus in this section is on use cases requiring local human interaction (e.g., video gaming or augmented reality), the focus in Section VI is on use cases requiring local interaction between objects (e.g., vehicles).

## VI. NATIVE SUPPORT FOR M2M COMMUNICATION

Wireless communication is becoming a commodity, just like electricity or water [13]. This commoditization, in turn, is giving rise to a large class of emerging services with new types of requirements. We point to a few representative such requirements, each exemplified by a typical service:

- **A massive number of connected devices**. Whereas current systems typically operate with, at most, a few hundred devices per base station, some M2M services might require over $10^4$ connected devices. Examples include metering, sensors, smart grid components, and other enablers of services targeting wide area coverage.
- **Very high link reliability**. Systems geared at critical control, safety, or production, have been dominated by wireline connectivity largely because wireless links did not offer the same degree of confidence. As these systems transition from wireline to wireless, it becomes necessary for the wireless link to be reliably operational virtually all the time.
- **Low latency and real-time operation**. This can be an even more stringent requirement than the ones above, as it demands that data be transferred reliably within a given time interval. A typical example is Vehicle-to-X connectivity, whereby traffic safety can be improved through the timely delivery of critical messages (e.g., alert and control).

Fig. 5 provides a perspective on the M2M requirements by plotting the data rate vs. the device population size. This cartoon illustrates where systems currently stand and how the research efforts are expanding them. The area R1 reflects the operating range of today's systems, outlining the fact that the device data rate decreases as its population increases. In turn, R2 is the region that reflects current research aimed at improving the spectral efficiency. Finally, R5 indicates the region where operation is not feasible due to fundamental physical and information-theoretical limits.

Regions R3 and R4 correspond to the emerging services discussed in this section:

- R3 refers to massive M2M communication where each connected machine or sensor transmits small data blocks sporadically. Current systems are not designed to simultaneously serve the aggregated traffic accrued from a large number of such devices. For instance, a current system could easily serve 5 devices at 2 Mbps each, but not 10000 devices each requiring 1 Kbps.
- R4 demarks the operation of systems that require high reliability and/or low latency, but with a relatively low average rate per device. The complete description of this region requires additional dimensions related to reliability and latency.

There are services that pose simultaneously more than one of the above requirements, but the common point is that *the data size of each individual transmission is small*, going down to several bytes. This profoundly changes the communication paradigm for the following reasons:

- Existing coding methods that rely on long codewords are not applicable to very short data blocks.
- Short data blocks also exacerbate the inefficiencies associated with control and channel estimation overheads. Currently, the control plane is robust but suboptimal as it represents only a modest fraction of the payload data; the most sophisticated signal processing is reserved for payload data transmission. An optimized design should aim at a much tighter coupling between the data and control planes.
- As mentioned in Section II, the architecture needs a major redesign, looking at new types of nodes. At a system level, the frame-based approaches that are at the heart of 4G need rethinking in order to meet the requirements for latency and flexible allocation of resources to a massive number of devices.

From the discussion above, and from the related architectural consideration in Section II, and referring one last time to the Henderson-Clark model, we conclude that a native support of M2M in 5G requires radical changes at both the node and the architecture level. Major research work remains to be done to come up with concrete and interworking solutions enabling 'M2M-inside' 5G systems.

## VII. CONCLUSION

This paper has discussed five disruptive research directions that could lead to fundamental changes in the design of cellular networks. We have focused on technologies that could lead to both architectural and component design changes: device-centric architectures, mmWave, massive-MIMO, smarter devices, and native support to M2M. It is likely that a suite of these solutions will form the basis of 5G.

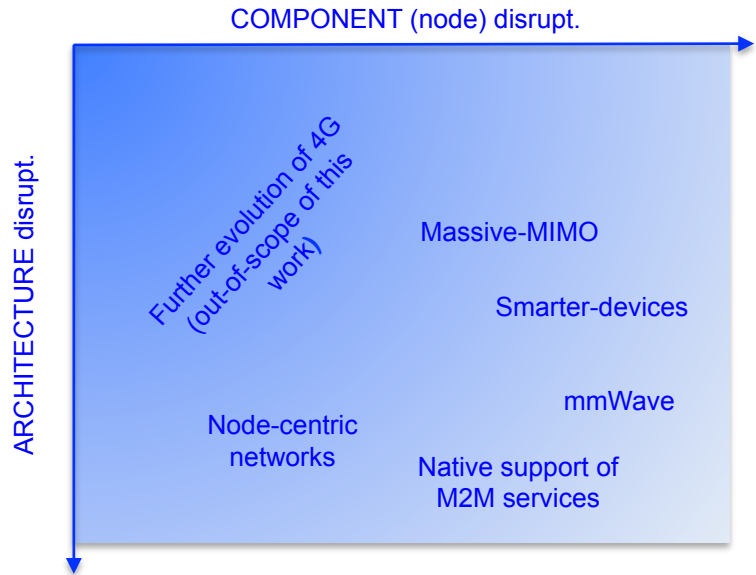

**Figure 1.** The five disruptive directions for 5G considered in this paper, classified according to the Henderson-Clark model.

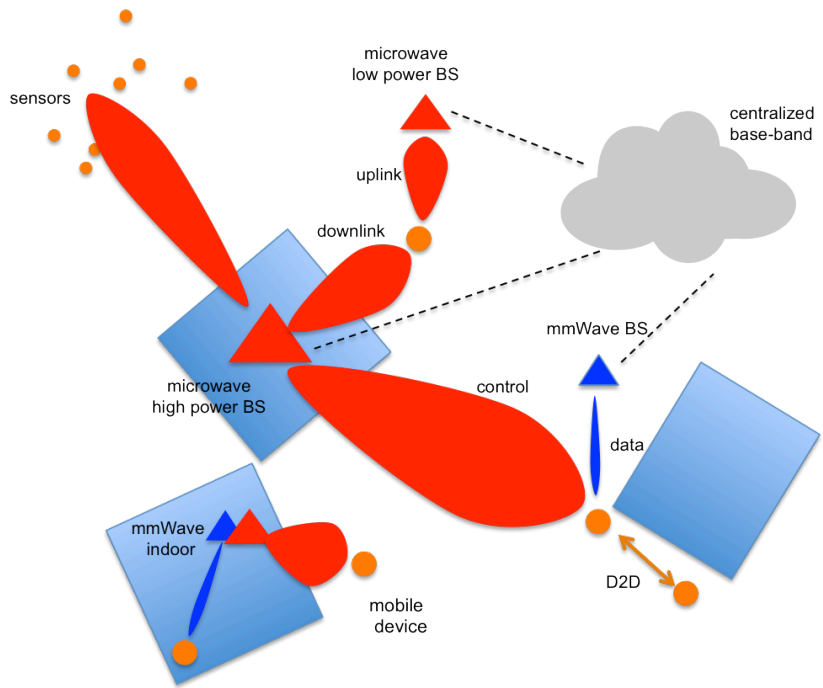

**Figure 2.** Example of device-centric architecture.

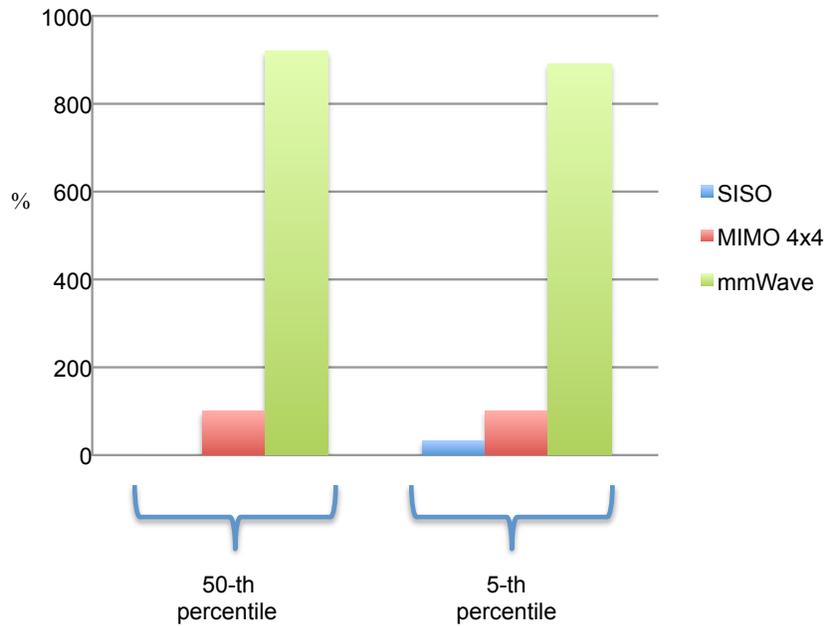

**Figure 3.** Cell data rate comparison between microwave systems using 50 MHz of bandwidth (single-user single-antenna and single-user MIMO) and a mmWave system with 500 MHz of bandwidth and a single user. Results are given in terms of gain (%) w.r..t. the MIMO 4x4 baseline (cf. footnote[3]). More details about the comparison setup are provided in [12].

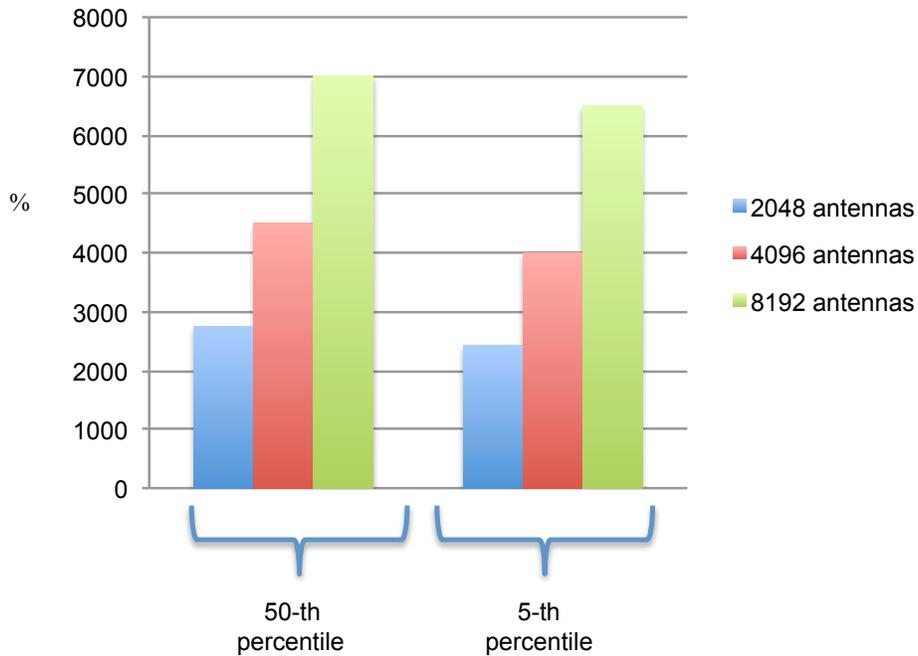

**Figure 4.** Cell data rate comparison for a fixed access application of massive-MIMO. An array of 2048, 4096 or 8192 antennas, utilizing 50 MHz and radiating a total of 120 Watts, serves 1000 users randomly located in a cell of radius 6 km. Results are given in terms of gain (%) w.r.t. the MIMO 4x4 baseline.[3]

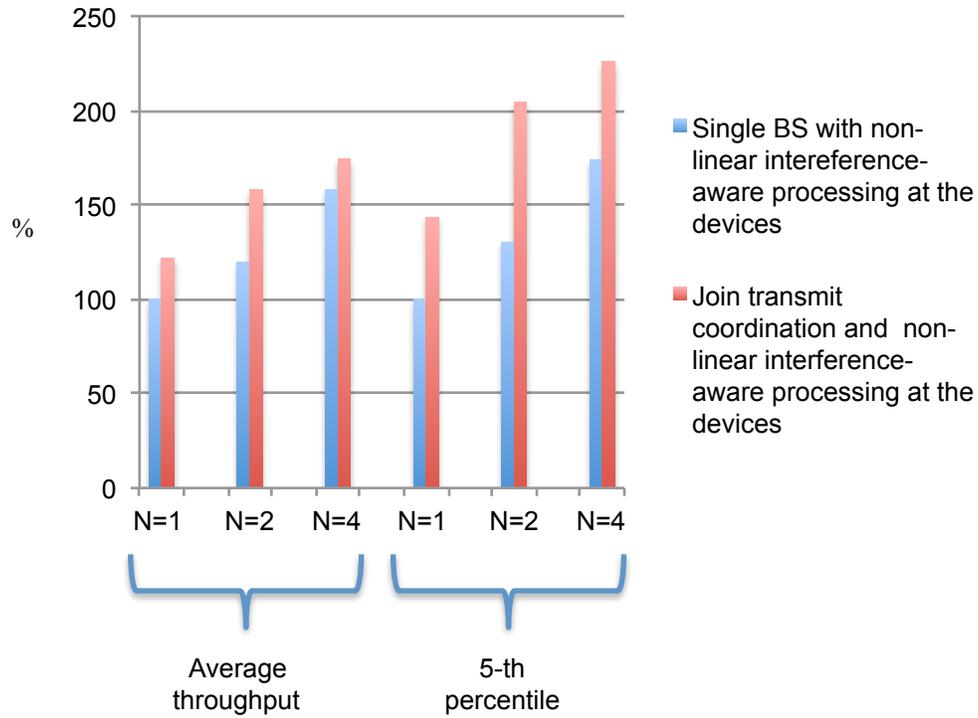

**Figure 5.** Throughput gains obtained by incorporating the effects of nonlinear, intra and inter-cluster interference-awareness into devices, with N = 1, 2 and 4 antennas. Results are given in terms of gain (%) w.r.t. the single-base single-antenna baseline. More details can be found in [16].

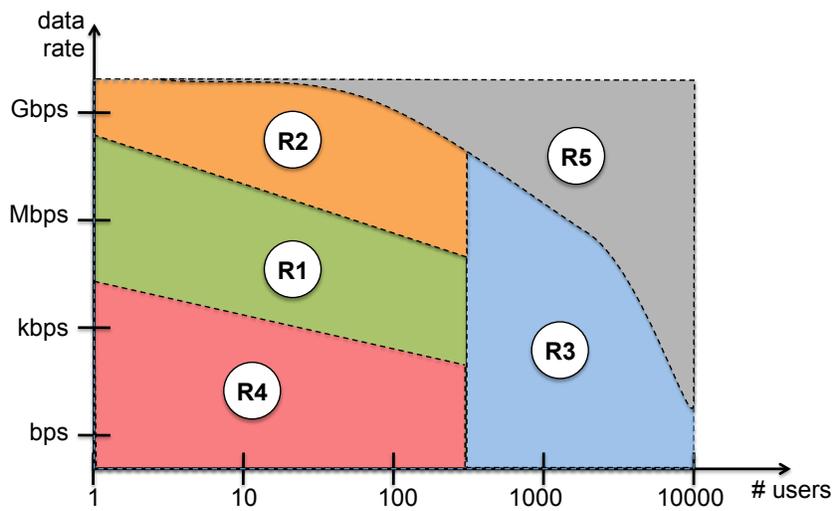

**Figure 6.** Operating regions in terms of data rate vs. size of the population.